\documentclass[a4paper,12pt]{article}

\begin{document}

\author{C. Barrab\`es\thanks{E-mail : barrabes@celfi.phys.univ-tours.fr}\\     
\small Laboratoire de Math\'ematiques et Physique Th\'eorique\\
\small  CNRS/UPRES-A, Universit\'e F. Rabelais, 37200 TOURS, France\\
P.A. Hogan\thanks{E-mail : phogan@ollamh.ucd.ie}\\
\small Mathematical Physics Department\\
\small  University College Dublin, Belfield, Dublin 4, Ireland}

\title{Light--Like Signals in General Relativity\\
and Cosmology}
\date{}
\maketitle

\begin{abstract}
The modelling of light--like signals in General Relativity 
taking the form of impulsive gravitational waves and 
light--like shells of matter is examined. Systematic deductions 
from the Bianchi identities are made. These are based upon Penrose's 
hierarchical classification of the geometry induced on 
the null hypersurface history of the signal by its 
embedding in the space--times to the future and 
to the past of it. The signals are not confined to propagate 
in a vacuum and thus their interaction with matter (a burst of 
radiation propagating through a cosmic fluid, for example) is 
also studied.  Results are accompanied by illustrative examples using 
cosmological models, vacuum space--times, the de Sitter 
universe and Minkowskian space--time. 
\end{abstract}
\thispagestyle{empty}
\newpage

\section{Introduction}\indent
In his classic work on impulsive gravitational waves (gravitational 
waves having a Dirac delta function profile) Penrose \cite{P} 
introduced a hierarchical classification of intrinsic geometries 
that the null hypersurface history of the wave front inherits 
from the space--times it is embedded in. This classification is 
related to the physical characteristics of the light--like signal -- 
a fact which emerges clearly from the analysis of light--like 
shells of matter and impulsive gravitational waves carried out 
by Barrab\`es and Israel \cite{BI}. They generalise the usual 
approach (see \cite{I}) to the study of non--null singular 
hypersurfaces, based on the extrinsic curvature tensor of the 
hypersurface, to include the light--like case. The light--like 
signal can be an impulsive gravitational wave or a light--like 
shell of matter or both. The latter situation could be viewed 
as a model of a burst of gravitational radiation accompanied by 
a burst of neutrinos from a supernova \cite{BBH}. In \cite{P} and 
\cite{BI} the properties of the signal which can be obtained 
from the Bianchi identities are either implicit in the work or 
are only partially explicitly derived. In this paper we take a 
systematic approach to the deductions one can make from the 
Bianchi identities depending upon the type of induced geometry 
(in Penrose's sense) on the history of the signal. As one 
moves through the hierarchy of geometries more information 
becomes available from the Bianchi identities. As our study 
is not confined to vacuum space--times we in addition 
examine the interaction of a light--like signal with matter (a 
burst of radiation propagating through the cosmic fluid). We 
therefore analyse in some detail the influence of the light--like 
signal on a congruence of time--like world--lines -- 
the histories of galaxies in a cosmological model, for example. 
We show that for the signal to include a gravitational wave 
the shear of the fluid lines must jump across the history 
of the signal in space--time. For the signal to be a 
light--like shell of matter then a jump in the acceleration, 
vorticity or expansion of the histories of the fluid lines is 
necessary. Our deductions are presented (in section 4 below) 
in the form of a series of Lemmas each followed by illustrative 
examples involving cosmological models, vacuum space--times, 
the de Sitter universe and Minkowskian space--time.

The outline of the paper is as follows: In section 2 we 
give an introduction to and brief summary of the Barrab\`es/Israel 
\cite{BI} formalism, the basic equations needed to study 
the behavior of a time--like congruence intersecting the history 
of the light--like signal in space--time and a description of the 
Penrose classification of induced geometries on the history 
of the signal. In section 3 we systematically deduce the 
consequences of the twice--contracted Bianchi identities and 
the Bianchi identities which become available as one works through the 
hierarchy of induced geometries. Section 4 contains the 
main conclusions of our work in the form of physical and 
geometrical applications of the results of sections 2 and 3. The 
paper concludes with a discussion in section 5.     
\setcounter{equation}{0}
\section{Geometrical Preliminaries}\indent
The history of a light--like shell or an impulsive 
gravitational wave 
corresponds in the space-time manifold $\cal M$ to a 
singular null hypersurface 
across which the metric tensor $g_{\mu \nu}$ is only $C^0$, 
i.e. it is continuous but its first derivatives are 
discontinuous across the hypersurface. This can be used for example as a model for 
the emission of light--like matter and gravitational radiation 
due to
a sudden change in properties such as the mass, the angular momentum 
and the mutipole moments of a gravitating body
(an example of this is the production of bursts of neutrinos 
and gravitational radiation from a supernova). In this section 
we first make a brief review of 
a general formalism adapted to the case of singular null 
hypersurfaces which
has been developed a few years ago \cite{BI}, 
and of its application to the splitting of the light--like signal 
between a shell and a wave \cite{BBH}. Then we analyze the 
discontinuities 
in the kinematical quantities (velocity, acceleration, expansion, shear and
vorticity) of a congruence of time--like lines crossing
a singular null hypersurface. Finally we present the Penrose classification
\cite{P} of intrinsic geometries that a singular null hypersurface 
inherits from its embedding in $\cal M$ and which will serve 
as a basis for the classification of the action of a null shell 
and/or a wave on a congruence
of timelike lines.

We consider a space-time $\cal M$ which is divided into two 
halves $\cal M^+$ and $\cal M^-$
separated by a null hypersurface $\cal N$ and with a local coordinate
system $\{x^{\mu}_+\}$ in $\cal M^+$ and 
$\{x^{\mu}_-\}$ in $\cal M^-$, Greek indices taking
values $0,1,2,3$, in terms of which the metric tensor components are 
$g^{+}_{\mu \nu}$ and $g^{-}_{\mu \nu}$ respectively. Let ${\xi}^a$,  
Latin indices taking values $1,2,3$, be local intrinsic coordinates 
on $\cal N$. The parametric equations of
the embeddings of $\cal N$ have the forms, 
$x^{\mu}_{\pm} = f^{\mu}_{\pm}({\xi}^a)$. The corresponding tangent basis vectors are 
$e_{(a)}=\partial / \partial {\xi}^a$, and 
their scalar products define the induced metric on $\cal N$:   
\begin{equation} \label{2.1}
  g_{ab} = e_{(a)}\cdot e_{(b)} = g_{\mu \nu}e^{\mu}_{(a)}\,
e^{\nu}_{(b)}\vert_{\pm}\ ,
\end{equation}
assumed the same on both faces of the hypersurface. Four dimensional 
scalar products are indicated by a dot as in (\ref{2.1}) and evaluation of a quantity 
on the plus or minus sides of $\cal N$ is indicated by a 
vertical stroke followed by a plus or minus (or both if the 
quantity is the same on both sides of $\cal N$).

Let $n$ be normal to $\cal N$ with components $n^{\mu}_{\pm}$ viewed on the
plus and minus sides. Thus
\begin{equation}
\label{2.2}
n\cdot n \vert_{\pm} = 0\ ,\hspace*{1cm} n\cdot e_{(a)} 
\vert_{\pm} = 0\ .
\end{equation}
As $\cal N$ is a null hypersurface the normal $n$ is tangent to it 
and can
be decomposed along the tangent basis vectors as
\begin{equation}
\label{2.3}
n = n^a e_{(a)}\ .
\end{equation} 
Let us introduce a transversal $N$ on $\cal N$ and require its
projections onto the plus and minus sides of $N$ to be equal, i.e.
\begin{equation}
\label{2.4}
[N\cdot N] = [N\cdot e_{(a)}] = 0\ ,
\end{equation}
where we use square brackets to denote the jump of any quantity $F$ that is
discontinuous across the hypersurface; $[F]={}^+F-{}^-F$.
In order to make sure that $N$ is transverse to the hypersurface we require
that it has a non-vanishing scalar product with the normal $n$;
\begin{equation}
\label{2.5}
N\cdot n = \eta^{-1}\ ,
\end{equation}
where $\eta$ is some non-zero constant usually taken to be $-1$. Note that
$N$ is not uniquely defined by this equation as one may 
make a tangential
displacement $N \mapsto N + \zeta^a ({\xi}^b) e_{(a)}$ with arbitrary functions
${\zeta}^a$ for a prescribed $\eta$.
Next we define the transverse extrinsic curvature ${\cal K}_{ab}$ of the
null hypersurface $\cal N$ as
\begin{equation}
\label{2.6}
{\cal K}_{ab} = -N\cdot \nabla _{e_{(b)}} e_{(a)}\ ,
\end{equation}
and its jump across the null hypersurface by
\begin{equation}
\label{2.7}
{\gamma}_{ab} = 2 [{\cal K}_{ab}]\ ,
\end{equation} 
where $\nabla$ denotes covariant differentiation with respect to 
the four-dimensional Levi--Civita connection. It can be shown \cite{BI} 
that the jumps $\gamma _{ab}$
are independent of the choice of the transversal $N$ and thus define
a purely intrinsic property of the hypersurface.

As the metric tensor is only $C^0$ at the hypersurface the Riemann, Ricci
and Weyl tensors in general each contain a singular Dirac 
$\delta$-term. The stress-energy
tensor which appears on the matter side of the Einstein field equations
also contains a similar term which one interprets as the surface stress-energy
tensor of a shell whose history in space-time is the null hypersurface 
$\cal N$. The existence of a surface stress-energy tensor is an 
intrinsic property 
of the hypersurface and it has intrinsic components 
$S^{ab}$ which are given by\cite {BI}
\begin{equation}
\label{2.8}
16 \pi {\eta}^{-1} S^{ab} = (g^{ac}_* n^b n^d + g^{bd}_* n^a n^c )\gamma _{cd}
- g^{ab}_* \gamma ^{\dag} - n^a n^b (g^{cd}_* \gamma _{cd})\ .
\end{equation}
where $\gamma^{\dag}=\gamma_{ab}n^a n^b$. In this expression 
we have introduced a (non unique) 
`pseudo'-inverse metric \cite {BI} $g^{ab}_*$ (as $\cal N$ is a null hypersurface the
induced metric is degenerate and has no inverse) such that
\begin{equation}
\label{2.9}
g^{ab}_* g_{bc} = \delta ^a_c - \eta n^a N_c\ ,
\end{equation}
where $N_c = N\cdot e_{(c)}$. It is important to note that 
the expression (\ref{2.8}) for the surface
stress-energy tensor is independent of the choice of the `pseudo'-inverse
metric.

Another important point is that only four of the six components of the jump
$\gamma _{ab}$ appear in $S^{ab}$, and correspond to 
$n^a \gamma _{ab}$ and $g^{ab}_* \gamma _{ab}$. This leaves two components
which are decoupled from matter and describe an impulsive gravitational
wave, as an analysis of the singular $\delta$-part of the Weyl tensor 
reveals \cite {BBH}.
These two components are given by \cite{BBH}
\begin{equation}
\label{2.1O}
\hat{\gamma}_{ab} = \gamma _{ab} - \frac{1}{2} g^{cd}_* 
\gamma_{cd}\,g_{ab} 
- 2 \eta\,n^c \gamma _{c(a} N_{b)} + \eta^2 \gamma_{cd} 
n^c n^d N_a N_b\ ,
\end{equation}
and it is easy to see that $\hat{\gamma}_{ab} n^b = 0$ and 
$g^{ab}_* \hat{\gamma}_{ab} = 0$. The two non-vanishing components 
of $\hat{\gamma}_{ab}$ represent the two degrees of 
freedom of polarization of the wave.

A covariant counterpart to the above description exists and we now
summarize the main results. The intrinsic quantities $\gamma_{ab}$ and $S^{ab}$
can be extended to four-tensors on the space-time manifold. In a local 
coordinate
system $\{x^{\mu}\}$ covering both sides of $\cal N$ (for example it can
be either $\{x^{\mu}_+\}$ or $\{x^{\mu}_-\}$) we define the tensors
$\gamma_{\mu \nu}$ and $S^{\mu \nu}$ by the requirements.
\begin{equation}
\label{2.11}
\gamma_{\mu \nu} e^{\mu}_{(a)} e^{\nu}_{(b)} = \gamma_{ab}\ , \hspace*{1cm} 
  S^{ab} e^{\mu}_{(a)} e^{\nu}_{(b)} = S^{\mu \nu}\ .
\end{equation}
It is easy to see that $S^{\mu \nu}$ satisfies $S^{\mu \nu}n_{\nu}=0$
and is thus tangential to the hypersurface.
It can also be shown that $\gamma_{\mu \nu}$ is directly related
to the jump in the first derivatives of the metric tensor. One gets
\begin{equation}
\label{2.12}
[\partial _{\alpha} g_{\mu \nu}] = \eta n_{\alpha} \gamma_{\mu \nu}\ 
,\hspace*{1cm} [N^{\alpha} \partial _{\alpha} g_{\mu \nu}] = 
\gamma_{\mu \nu}\ ,
\end{equation}
where we have used (\ref{2.5}), and for the jumps of the Christoffel
symbols
\begin{equation}
\label{2.13}
[\Gamma^{\mu}_{\alpha \beta}] = \frac{\eta}{2}(\gamma^{\mu}_{\alpha}n_{\beta}
+\gamma^{\mu}_{\beta}n_{\alpha} - n^{\mu}\gamma_{\alpha \beta})\ .
\end{equation}
The covariant components $S_{\alpha \beta}$ of the surface 
stress-energy tensor of the null shell are given by
\begin{equation}
\label{2.14}
16 \pi \eta^{-1} S_{\alpha \beta} = 2 \gamma_{(\alpha}n_{\beta)} 
-\gamma n_{\alpha}n_{\beta} -\gamma^{\dag} g_{\alpha \beta}\ ,
\end{equation}
where
\begin{equation}
\label{2.15}
\gamma_{\alpha} = \gamma_{\alpha \beta}n^{\beta}\ ,\hspace*{1cm} 
\gamma = \gamma_{\alpha \beta}g^{\alpha \beta}\ ,
\end{equation}
and $\gamma^{\dag}$ is defined above and can be rewritten as
\begin{equation}
\label{2.16}
\gamma^{\dag} = \gamma_{\alpha \beta}n^{\alpha}n^{\beta} = 
\gamma_{\alpha}n^{\alpha}\ .
\end{equation}

We consider now a congruence of timelike curves with unit tangent
vector $u^{\alpha}$, $u\cdot u=-1$, crossing the singular null hypersurface
$\cal N$ (as previously we are using a coordinate system $\{x^{\mu}\}$ 
covering both sides of $\cal N$). This congruence can be arbitrarily 
chosen in each domain $\cal M^{\pm}$
and in the sequel we shall only consider the case where the tangent
vector is continuous across $\cal N$. If $u$ is tangent to 
matter world--lines in $\cal M^{\pm}$ then choosing $u$ 
continuous across $\cal N$ forbids $\cal N$ becoming the history 
of a shock wave propagating through the matter in the usual 
sense (\cite{WI}, \cite{RG}). We choose $u$ to be continuous across 
$\cal N$ for the following reasons: (a) this is the minimal 
requirement consistent with a delta function appearing in the 
Weyl tensor (as can be seen by applying the Ricci identities to 
$u$), (b) this allows finite jumps (and no delta function) to 
appear in the kinematical quantities associated with the integral 
curves of $u$ (the acceleration, vorticity, shear and 
expansion) and (c) the jumps in the kinematical quantities are then 
simply related to the presence or otherwise of a light--like shell 
and /or a gravitational wave in the signal with history $\cal N$ 
(see section 4 below where these general results and physical 
examples are given).
 
Let us then consider a congruence of timelike curves with continuous
4--velocity $u$ but with discontinuous first derivatives, 
having jumps across $\cal N$ described by the vector $\lambda$ such that
\begin{equation}
\label{2.17}
[\partial _{\mu} u^{\alpha}] = \eta n_{\mu} \lambda^{\alpha} 
\hspace*{0.5cm}{\rm or} \hspace*{0.5cm} [N^{\mu} 
\partial _{\mu} u^{\alpha}] = \lambda^{\alpha}\ ,
\end{equation}
where we have used (\ref{2.5}). It follows that the jump in the 4--acceleration
$a^{\alpha}=u^{\mu} \nabla _{\mu}u^{\alpha}$ is given by
\begin{equation}
\label{2.18}
\eta ^{-1} [a^{\alpha}] = -s \lambda^{\alpha} -s U^{\alpha} -
\frac{1}{2}(u\cdot U)n^{\alpha}\ ,
\end{equation}
where we have put $s=-u\cdot n>0$ and $U_{\alpha}=\gamma _{\alpha \beta} u^{\beta}$.
Using $u\cdot a=0$ and the above equations one gets the following relation
\begin{equation}
\label{2.19}
u\cdot U = -2u\cdot \lambda\ .
\end{equation}
The expansion $\theta$, shear $\sigma _{\alpha \beta}$, and 
vorticity 
$\omega _{\alpha \beta}$ of the timelike congruence are in general 
discontinuous across $\cal N$ and have jumps given by
\begin{equation}
\label{2.20}
\eta ^{-1}[\theta] = - \frac{1}{2}s \gamma + \lambda\cdot n\ ,
\end{equation}
\begin{eqnarray}
\label{2.21}
\eta^{-1} [\sigma_{\alpha \beta}] & = & - \frac{s}{2}\gamma_{\alpha \beta} + 
\lambda_{(\alpha}n_{\beta)} - sU_{(\alpha}u_{\beta)} - 
\frac{1}{2}(u\cdot U)
n_{(\alpha}u_{\beta)}{}\nonumber\\
& & {}- s\lambda_{(\alpha}u_{\beta)} - 
\frac{\eta^{-1}}{3}
[\theta]h_{\alpha \beta}{}\ ,
\end{eqnarray}
\begin{equation}
\label{2.22}
\eta^{-1}[\omega_{\alpha \beta}]=U_{[\alpha}n_{\beta]} + 
\lambda_{[\alpha}n_{\beta]} - sU_{[\alpha}u_{\beta]}-
\frac{1}{2}(u\cdot U)n_{[\alpha}u_{\beta]} 
-s\lambda _{[\alpha} u_{\beta]}\ ,
\end{equation}
where the projection tensor 
$h_{\alpha \beta}=g_{\alpha \beta} + u_{\alpha}u_{\beta}$
has been introduced. We can alternatively express (\ref{2.22}) 
in terms of the vorticity vector
\begin{equation}\label{2.22'}
\omega ^{\alpha}=\frac{1}{2}\,\eta ^{\alpha\beta\mu\nu}\,
u_{\beta}\,\omega _{\mu\nu}\ ,
\end{equation}
where $\eta ^{\alpha\beta\mu\nu}=(-g)^{-\frac{1}{2}}\epsilon 
_{\alpha\beta\mu\nu}$ with $g={\rm det} (g_{\mu\nu})$ and 
$\epsilon _{\alpha\beta\mu\nu}$ is the four--dimensional Levi--
Civita permutation symbol. The jump in the vorticity vector 
is given by
\begin{equation}\label{2.22*}
\eta ^{-1}\left [\omega ^{\alpha}\right ]=\frac{1}{2}\,
\eta ^{\alpha\beta\mu\nu}\,u_{\beta}\,\left (U_{\mu}+
\lambda _{\mu}\right )\,n_{\nu}\ .
\end{equation} 

Finally we recall a classification introduced by Penrose \cite{P}
for the geometry that the null hypersurface $\cal N$ inherits 
from $\cal M^{+}$ and $\cal M^{-}$. As any
null hypersurface $\cal N$ is generated by a uniquely defined 
two-parameter family of null geodesics one can consider a hierarchy of 
three types of intrinsic geometry in order of increasing structure\\
-Type I: the induced metrics match (our basic assumption 
(\ref{2.1})),\\
-Type II: Type I with parallel transport of the normal $n$ along the null generators
matching,\\
-Type III: Type II with parallel transport of any tangent vector 
to $\cal N$ along the
null generators matching.\\
A Type I geometry on $\cal N$ is the most general of the 
three types and is always 
assumed to be valid in our considerations. 
A Type II geometry requires that 
$n^{\mu}\nabla^{+}_{\mu}n^{\alpha} = n^{\mu}\nabla^{-}_{\mu}n^{\alpha}$
or using (\ref{2.13}) that $\gamma^{\dag} = 0$. If one defines the 
acceleration parameter $\kappa$ by
\begin{equation}
\label{2.23}
n^{\mu}\nabla_{\mu}n^{\alpha} = \kappa n^{\alpha}\ ,
\end{equation} one can show that $[\kappa]=\eta\,\gamma^{\dag}/2$ so that
a Type II geometry implies $[\kappa]=0$. In particular this is realized
when the null generators are affinely parametrized on both sides 
of $\cal N$. A Type III geometry requires that
$n^{\mu}\nabla^{+}_{\mu}v^{\alpha} = n^{\mu}\nabla^{-}_{\mu}v^{\alpha}$ 
for any vector $v^{\alpha}$ such that $n_{\alpha}v^{\alpha}=0$, 
and using (\ref{2.13}) this implies that $\gamma_{\alpha} = A n_{\alpha}$
where $A$ is an arbitrary function on $\cal N$.
 
A physical interpretation in terms of the surface stress-energy tensor
of a null shell, if it exists having history $\cal N$, 
can be given to the Type II and III 
geometries above. As shown in \cite{BI}
the surface pressure $p= -(\eta /16\pi)\gamma^{\dag}$ and so a Type II 
induced geometry corresponds to a pressure-free null shell, i.e. a shell only 
admitting a surface energy density $\eta\,f$ and anisotropic 
surface stresses. For a Type III geometry induced on $\cal N$ 
the surface stress-energy
tensor reduces from (\ref{2.14}) to 
$16 \pi \eta^{-1 }S_{\alpha \beta} = (2A-\gamma) 
n_{\alpha}n_{\beta}$, 
there are no surface stresses and the surface energy density is 
$\eta\,f=\eta\,(2A-\gamma)/16\pi$.
If $A=\gamma/2$ then there is no shell and $\cal N$ is the 
history of an impulsive gravitational wave provided $\hat\gamma _{ab}\neq 0$.
\setcounter{equation}{0}
\section{Consequences of the Bianchi Identities}\indent
On the null hypersurface history of the light--like signal we 
have the normal $n$ (which is tangential to $\cal N$ since $\cal N$ 
is null) and the 
time--like vector field $u$, for which $u\cdot n=-s<0$. In a local 
coordinate system $\{x^\mu\}$ covering both sides of $\cal N$, 
it is helpful to define on $\cal N$, but not tangent to $\cal N$, 
the null vector field
\begin{equation}
\label{3.1}
l^{\mu}=-\frac{1}{2s^2}\,n^{\mu}+\frac{1}{s}\,u^{\mu}\ ,
\end{equation}with $l\cdot n=-1$. Now $n,\ l$ can be supplemented 
by a complex null vector field $m$, with components $m^{\mu}$, 
tangent to $\cal N$ and also orthogonal to $l$, and satisfying 
$m\cdot \bar m=+1$ (the bar denoting complex conjugation). 
The null tetrad $\{n, l, m, \bar m\}$ defined on $\cal N$ will be 
useful for the purposes of displaying formulas below. If $\cal N$ 
has equation $\Phi (x^\mu )=0$ and $n_\mu =\alpha ^{-1}\Phi _{,\mu}$ 
for some function $\alpha$ defined on $\cal N$ then following from 
the results of section 2, the components of the Einstein tensor 
of the space--time $\cal M^+\cup\cal M^-$ have the form 
\begin{equation}\label{3.2}
G^{\mu\nu}=\hat G^{\mu\nu}\,\delta\left (\Phi\right ) +\Theta
\left (\Phi\right )\,{}^+G^{\mu\nu}+
\left (1-\Theta\left (\Phi\right)\right )\,{}^-G^{\mu\nu}\ ,\end{equation}
where $\delta$ is the Dirac delta function, $\Theta$ is the 
Heaviside step function with $\Theta >0$ in $\cal M^+$, $\Theta <0$ in 
$\cal M^-$ and
\begin{equation}\label{3.3}
\Theta _{,\mu}=\alpha\,n_{\mu}\,\delta\left (\Phi\right )\ .\end{equation}

In (\ref{3.3}) ${}^{\pm}G^{\mu\nu}$ are the components of the 
Einstein tensors in $\cal M^{\pm}$ respectively and can be 
written as $-8\pi\,{}^{\pm}T^{\mu\nu}$ in terms of the 
respective energy--momentum--stress tensors. Also
\begin{equation}\label{3.4} 
\hat G^{\mu\nu}=-8\pi\,\alpha\,S^{\mu\nu}\ ,\end{equation}
with $S^{\mu\nu}$ given by (\ref{2.14}). Thus in particular 
\begin{equation}\label{3.5}
\hat G^{\mu\nu}\,n_\nu =0\ .\end{equation}
We now apply the twice--contracted Bianchi identities 
$\nabla _{\nu}G^{\mu\nu}\equiv 0$ to (\ref{3.2}). On 
account of (\ref{3.5}) the term in $\nabla _{\nu}G^{\mu\nu}$ 
involving the derivative of the delta function vanishes and 
we obtain
\begin{equation}\label{3.5'}
\nabla _{\nu}\hat G^{\mu\nu}+\alpha\,\left [G^{\mu\nu}\,n_\nu\right ]
=0.\end{equation} 
Since $\hat G^{\mu\nu}$ in (\ref{3.2}) is defined $\it on$ 
$\cal N$ it only makes sense to calculate derivatives of 
$\hat G^{\mu\nu}$ tangential to $\cal N$. The number of 
such different tangential derivatives available to us 
depends upon the type of the induced geometry on $\cal N$. 
If the induced geometry is Type I then we obtain 
one meaningful equation from (\ref{3.5'}), 
namely, 
\begin{equation}\label{3.6}
-8\pi\,\eta ^{-1}\left [T_{\mu\nu}\,n^\mu\,n^\nu\right ]=\rho\,
\gamma ^{\dagger}\ .\end{equation}
Here square brackets as always denote the jump in the 
enclosed quantity across $\cal N$ and $\rho =m^\mu\,\bar m^\nu\,
\nabla _{\nu}n_\mu$ is the expansion of the null geodesic 
integral curves of $n$ ($[\rho ]=0$ since $\rho$ is intrinsic 
to $\cal N$). If the induced geometry is Type II then 
$\gamma ^{\dagger}=0$. Hence the acceleration parameter $\kappa$ 
introduced in (\ref{2.23}) is continuous across $\cal N$. 
Now the meaningful equations emerging from the 
twice contracted Bianchi identities are (\ref{3.6}) with 
$\gamma ^{\dagger}=0$ and also 
\begin{equation}\label{3.7}-16\pi\,\eta ^{-1}\left [T_{\mu\nu}\,m^\mu\,n^\nu\right ]=
\gamma '_{\mu}\,m^\mu +3\rho\,\gamma _{\mu}\,m^\mu +\sigma\,\bar 
m^\mu\,\gamma _\mu\ .\end{equation}
Here $\gamma '_{\mu}=n^\nu\,\nabla _{\nu}\,\gamma _\mu$ and 
$\sigma =m^\mu\,m^\nu\,\nabla _{\nu}n_{\mu}$ is the complex 
shear of the generators of $\cal N$, which 
is intrinsic even for a Type I geometry. Finally if the 
induced geometry is Type III then $\gamma _\mu =A\,n_\mu$ 
for some function $A$ defined on $\cal N$. Now the right hand 
sides of (\ref{3.6}) and (\ref{3.7}) vanish and in addition 
we find that 
\begin{equation}\label{3.8}
\eta ^{-1}\left [T_{\mu\nu}\,l^\mu\,n^\nu\right ]=f'+
(2\rho +\kappa)\,f\ ,\end{equation}
where (as in section 2) $8\pi\,f=A-\frac{1}{2}\gamma$ and $f'=f_{,\mu}\,n^\mu$.

To obtain the Bianchi identities we use the tensor 
representing the left and right duals of the Riemann 
curvature tensor \cite{RR}
\begin{equation}\label{3.9}
G^{\mu\nu\rho\sigma}=\frac{1}{4}\,\eta ^{\mu\nu\alpha\beta}\,\eta ^{
\rho\sigma\lambda\gamma}\,R_{\alpha\beta\lambda\gamma}\ ,
\end{equation}
where $R_{\alpha\beta\lambda\gamma}$ are the components of 
the Riemann curvature tensor and $\eta ^{\alpha\beta\gamma\delta}=
(-g)^{-\frac{1}{2}}\,\epsilon _{\alpha\beta\gamma\delta}$. 
Then the Bianchi identities read 
\begin{equation}\label{3.10}
\nabla _{\sigma}\,G^{\mu\nu\rho\sigma}\equiv 0\ .\end{equation}
The right hand side of (\ref{3.9}) can be written 
in terms of the Riemann and Ricci tensors and the Ricci scalar as 
\begin{eqnarray}\label{3.11}
-G^{\mu\nu\rho\sigma}&=&R^{\mu\nu\rho\sigma}-g^{\mu\rho}\,R^{\nu\sigma}
-g^{\nu\sigma}\,R^{\mu\rho} {}\nonumber\\
& & {}+g^{\mu\sigma}\,R^{\nu\rho}
+g^{\nu\rho}\,R^{\mu\sigma}+\frac{1}{2}\,R\left (g^{\mu\rho}\,g^{\nu\sigma}
-g^{\mu\sigma}\,g^{\nu\rho}\right ){}\ .\end{eqnarray}
It is then useful to substitute in (\ref{3.11}) for the 
Riemann tensor in terms of the Weyl tensor. For the space--time 
$\cal M^+\cup\cal M^-$ the tensor (\ref{3.9}) has a 
decomposition similar to that of the Einstein tensor (\ref{3.2})
\begin{equation}\label{3.12}
G^{\mu\nu\rho\sigma}=\hat G^{\mu\nu\rho\sigma}\,\delta\left (\Phi\right ) +
\Theta\left (\Phi\right )\,{}^+G^{\mu\nu\rho\sigma}+\left (
1-\Theta\left (\Phi\right )\right )\,
{}^-G^{\mu\nu\rho\sigma}\ ,\end{equation}
with
\begin{equation}\label{3.13}
\eta ^{-1}\alpha ^{-1}\,\hat G^{\mu\nu\rho\sigma}=-2n^{[\mu}\,\gamma ^{\nu ]
\,[\rho}\,n^{\sigma ]}-2g^{\mu [\sigma}\,w^{\rho ]\nu}+
2\,g^{\nu [\sigma}\,w^{\rho ]\mu}-\gamma ^{\dagger}\,g^{\mu [\rho}\,g^{\sigma ]\nu}\ ,\end{equation}
where $w^{\mu\nu}=\gamma ^{(\mu}\,n^{\nu )}-\frac{1}{2}\,\gamma\,
n^\mu\,n^\nu$. One readily sees that
\begin{equation}\label{3.14}
\hat G^{\mu\nu\rho\sigma}\,n_{\sigma}=0\ .\end{equation}
Thus when (\ref{3.12}) is applied to (\ref{3.10}) the 
term involving the derivative of the delta function vanishes and we 
obtain from (\ref{3.10})
\begin{equation}\label{3.15}
\nabla _{\sigma}\,\hat G^{\mu\nu\rho\sigma}+\alpha\,\left [G^{\mu\nu\rho\sigma}\,
n_\sigma \right]=0\ .\end{equation}
Now $\hat G^{\mu\nu\rho\sigma}$ is defined on $\cal N$ and 
so (\ref{3.15}) only makes sense when it involves derivatives of 
$\hat G^{\mu\nu\rho\sigma}$ tangential to $\cal N$. This depends on the type of geometry 
induced on $\cal N$ by its embedding in $\cal M^+$ and in 
$\cal M^-$. If the geometry is Type I then we conclude from 
(\ref{3.15}):
\begin{equation}\label{3.16}
-8\pi\,\eta ^{-1}\left [T_{\mu\nu}\,n^\mu\,n^\nu\right ]=
\rho\,\gamma ^{\dagger}\ ,\end{equation}
\begin{equation}\label{3.17}
2\,\eta ^{-1}\left [\Psi _0\right ]=-\sigma\,\gamma ^{\dagger}\ ,\end{equation}
\begin{equation}\label{3.18}
2\,\eta ^{-1}\left [\Psi _{1}\right ]-8\pi\,\eta ^{-1}
\left [T_{\mu\nu}m^\mu\,n^\nu\right ]
=\rho\,m^\mu\gamma _\mu-\sigma\,\bar m^\mu\gamma _\mu\ ,
\end{equation}
Here (\ref{3.16}) coincides with the equation (\ref{3.6}) 
obtained from the twice--contracted Bianchi identities in the 
Type I case. $\left [\Psi _0\right ]$ and $\left [\Psi _1\right ]$ 
are the jumps in the Newman--Penrose components of the Weyl tensors 
of $\cal M^+$ and $\cal M^-$ across $\cal N$ (we use a standard 
notation $\cite{ES}$ for the components of the Weyl tensor, 
calculated on either side of $\cal N$, on the null tetrad 
$\{n, l, m, \bar m\}$). If the geometry induced on $\cal N$ is 
Type II then  (\ref{3.16}-\ref{3.18}) hold, with zeros on the right hand 
sides of (\ref{3.16}, \ref{3.17}), and we have, from (\ref{3.15}),
\begin{equation}\label{3.19}
2\,\eta ^{-1}\left [\Psi _1\right ]+8\pi\,\eta ^{-1}\left [T_{\mu\nu}\,
m^\mu\,n^\nu\right ]=-2\,\sigma\,\bar m^\mu\,\gamma _\mu -2\,
\rho \,m^\mu\,\gamma _\mu-m^\mu\,\gamma '_{\mu}\ .
\end{equation}
We note that the difference of equations (\ref{3.19}) and 
(\ref{3.18}) yields the equation (\ref{3.7}) obtained already 
from the twice--contracted Bianchi identities in the Type II case. 
Finally if the geometry of $\cal N$ is Type III then (\ref{3.16}-
\ref{3.19}) hold with the right hand sides all vanishing and in 
addition we have
\begin{equation}\label{3.20}
\eta ^{-1}\left [\Psi _2\right ]-\frac{2\pi}{3}\,\eta ^{-1}\left [T\right ]=
-\frac{1}{2}\,\sigma\,\gamma _{\mu\nu}\,\bar m^\mu\,\bar m^\nu +4\pi\,
\{f'+(\rho +\kappa )\,f\}\ ,\end{equation}
\begin{equation}\label{3.21}
-4\pi\,\eta ^{-1}\left [T_{\mu\nu}\,\bar m^\mu\,\bar m^\nu\right ]=
\frac{1}{2}\,\gamma '_{\mu\nu}\bar m^\mu\,\bar m^\nu+\frac{1}{2}\,
(\rho +\kappa )\,\gamma _{\mu\nu}\,\bar m^\mu\,\bar 
m^\nu -4\pi\,f\,\bar\sigma\ ,\end{equation}
where $\gamma '_{\mu\nu}=n^\sigma\,\nabla _\sigma \gamma _{\mu\nu}$.

As a final preliminary we note that if the coefficient of 
the delta function in the Weyl tensor $\hat C^{\mu\nu\rho\sigma}$ is 
calculated using $\hat G^{\mu\nu\rho\sigma}$ given by (\ref{3.13}) and $S^{\mu
\nu}$ given by (\ref{2.14}) and then if its components are 
calculated in the Newman--Penrose form $\hat\Psi _A$ (say) for 
$A=0, 1, 2, 3, 4$ we obtain 
\begin{equation}\label{3.22}
\hat\Psi _0=0,\ \hat\Psi _1=0,\ \hat\Psi _2=-\frac{1}{6}\,\eta\,\gamma 
^{\dagger},\ \hat\Psi _3=-\frac{1}{2}\,\eta\,\gamma _\mu\,\bar m^\mu ,\ 
\hat\Psi _4=-\frac{1}{2}\,\eta\,\gamma _{\mu\nu}\,\bar 
m^\mu\,\bar m^\nu\ .\end{equation}
This shows (cf. \cite{P}, \cite{BI}) that the delta function 
in the Weyl tensor is in general Petrov Type II. If the 
induced geometry on $\cal N$ is Type II then $\gamma ^{\dagger}
=0$ and $\hat\Psi _A$ is Petrov Type III whereas if the 
induced geometry on $\cal N$ is Type III then $\gamma ^{\dagger}
=0$ and $\gamma _{\mu}\,\bar m^\mu =0$ and $\hat\Psi _A$ is 
Petrov Type N. The signal with history $\cal N$ contains a gravitational 
wave if $\hat\Psi _4\neq 0$.
\setcounter{equation}{0}
\section{Physical and Geometrical Applications}\indent
We now draw physical and geometrical conclusions from the 
results outlined in sections 2 and 3, in the form of a series 
of Lemmas with illustrative examples.

Combining the jumps (\ref{2.18}), (\ref{2.20})--(\ref{2.22}) 
across $\cal N$, in the kinematical quantities associated 
with a time--like congruence intersecting $\cal N$, with 
the Newman--Penrose components $\hat\Psi _A$ (given by 
(\ref{3.22})) of the coefficients of the $\delta$--function in the 
Weyl tensor, we obtain, with straightforward algebra:

\vskip 1truepc

\noindent
Lemma 1: 

 (1) $\left [\sigma _{\mu\nu}\,m^\mu\,m^\nu\right ]\neq 0\ 
\Leftrightarrow\ 
\hat\Psi _4\neq 0$\ ;\\

 (2) {\it If}\hspace{0.25cm}$[\sigma _{\mu\nu}]=0\hspace*{0.25cm}{\it then}
\hspace*{0.25cm}\hat\Psi _4=0\hspace*{0.25cm} {\it and}$:\\

\hspace{0.5cm}(a)\hspace{0.25cm}$[a^\mu\,m_\mu]\neq 0\ \Leftrightarrow
\ \hat\Psi _3\neq 0$\ ;\\

\hspace{0.5cm}(b)\hspace{0.25cm}$[\omega ^\mu]\neq 0\ \Leftrightarrow\ 
\hat\Psi _3\neq 0$\ ;\\

 (3) {\it If}\hspace{0.25cm}$[\sigma _{\mu\nu}]=0\hspace{0.25cm}{\it and}
\hspace{0.25cm}[a^\mu]=0\hspace{0.25cm}
{\it then}\hspace{0.25cm}\hat\Psi _3=\hat\Psi _4=0\hspace{0.25cm}{\it and}
\hspace{0.25cm}[\theta ]\neq 0\ \Leftrightarrow\ 
\hat\Psi _2\neq 0$\ .

\vspace{0.5cm}

\noindent
We note from (2.18) and (2.24) that $[a^{\mu}]=0\ \Rightarrow\ 
[\omega ^\mu ]=0$. The converse is not true because again from 
(2.18) and (2.24) we find that if $[\omega ^\mu ]=0$ then $[a^\mu]=
s^{-2}n_{\lambda}\,[a^{\lambda}]\,(n^{\mu}-s\,u^{\mu})$ and we 
can only conclude from this that in general $m_{\mu}[a^{\mu}]=
0$. This explains the appearance of all components of $[\omega ^\mu ]$ 
in part (2b) of the Lemma and of only one complex component of 
$[a^\mu ]$ in part (2a) of the Lemma.
  
We are particularly interested in Lemma 1 when the time-like 
congruence of integral curves of $u$ are the world--lines of 
the cosmic fluid in a cosmological model. The first part of the 
Lemma says that if the signal with history $\cal N$ includes a 
gravitational wave then its effect on the cosmic fluid is 
to cause a jump across $\cal N$ in a complex component of 
the shear of the congruence and, if the passage of the signal 
through the fluid does not result in a jump in the fluid 
shear then the signal cannot contain a gravitational wave. In 
this latter case the signal is a light--like shell of matter 
with a Petrov Type II delta function in the Weyl tensor if 
the vorticity of the fluid jumps across $\cal N$ or if a 
complex component of the fluid 4--acceleration jumps across 
$\cal N$. If only the expansion of the fluid jumps across 
$\cal N$ then part (3) of the Lemma shows that the delta 
function in the Weyl tensor is Petrov Type III. 
 
There is an interesting analogy between Lemma 1 and the usual 
decomposition of perturbations of cosmological models into 
scalar, vector and tensor parts with the tensor perturbations 
describing propagating gravitational waves and the other 
perturbations describing inhomogeneity in the matter 
distribution (see \cite{EB} and \cite{HE}, for example). In Lemma 1 
the analogue of the tensor perturbations is the jump in the 
shear of the time--like congruence which by part (1) is 
necessary for the signal with history $\cal N$ to include 
a gravitational impulse wave. The analogues of the 
vector and scalar perturbations are the jumps in the 4--acceleration 
and vorticity on the one hand and in the expansion on the 
other hand leading, by parts (2) -- (3), 
to the possibility of the signal being a light--like shell 
of matter.

To illustrate Lemma 1 with an example of a signal consisting of 
a gravitational impulsive wave and a light--like shell propagating 
through the Einstein--de Sitter universe (say) we must choose 
a cosmological model left behind by the signal (the space--time 
$\cal M^+$ to the future of the null hypersurface $\cal N$) which 
has the properties: (a) its fluid 4--velocity joins continuously 
to that of the Einstein--de Sitter on $\cal N$ and (b) its 
fluid 4--velocity has shear. Thus the line--element of $\cal M^-$ 
is that of Einstein--de Sitter which, in 
coordinates $x_-^\mu=(t, r, \phi , z)$, reads
\begin{equation}\label{4.1}
ds^2=-dt^2+t^{4/3\beta}\,(dr^2+r^2d\phi ^2+dz^2)\ ,
\end{equation}
where $\beta$ is a constant. Here the $t$--lines are the world--
lines of the particles of a perfect fluid with isotropic pressure $p$ and 
proper--density $\mu$ satisfying the equation of state $p=(
\beta -1)\,\mu$. A simple example of a space--time $\cal M^+$ 
satisfying the requirements (a) and (b) above is the anisotropic Bianchi I 
space--time \cite{VE} with line--element, in coordinates 
$x^\mu _+=(t_+, r_+, \phi _+, z_+)$, 
\begin{equation}\label{4.2}
ds_+^2=-dt_+^2+A_+^2(dr_+^2+r^2_+d\phi _+^2)+B_+^2dz_+^2\ ,
\end{equation}
where
\begin{equation}\label{4.3}
A_+=t_+^{(3\beta -2)/6\beta}\ ,\qquad B_+=t_+^{2/3\beta}\ .
\end{equation} 
The $t_+$--lines are the world--lines of a perfect fluid with 
isotropic pressure $p_+$ and proper--density $\mu _+$ satisfying 
$p_+=\mu _+$. As boundary between $\cal M^-$ and $\cal M^+$ 
take the null hypersurface $\cal N$ to be given by
\begin{equation}\label{4.4}
r_+=T_+(t_+)\ ,\qquad \frac{dT_+}{dt_+}=\frac{1}{A_+}\ ,
\end{equation}
in the plus coordinates and by 
\begin{equation}\label{4.5}
r=T(t)\ ,\qquad \frac{dT}{dt}=t^{-2/3\beta}\ ,
\end{equation}
in the minus coordinates. As intrinsic coordinates on $\cal N$ we 
can use $\xi ^a=(r, \phi , z)$. The induced line--elements on $\cal N$ 
from $\cal M^+$ and $\cal M^-$ match (as required by (\ref{2.1})) 
if
\begin{equation}\label{4.6}
t_+=t,\ r_+=\frac{6\beta}{(3\beta +2)}\left [\frac{(3\beta -2)}{3\beta}
r\right ]^{\frac{(3\beta +2)}{(6\beta -4)}},\ \phi _+=\left (\frac{3\beta +2}
{6\beta -4}\right )
\phi ,\ z_+=z\ .
\end{equation}
We must first check that the 4--velocities of the fluid particles 
with histories in $\cal M^+$ and $\cal M^-$ are continuous across 
$\cal N$. This has to be done with care as we now have two local coordinate 
systems $\{x^\mu _+\}$ and $\{x^\mu _-\}$ on either side of $\cal N$, 
overlapping on $\cal N$ according to (\ref{4.6}). Let ${}^+v^\mu =(1, 
0, 0, 0)$ and ${}^-u^\mu =(1, 0, 0, 0)$. Then ${}^+v^\mu ,\ {}^-u^\mu$ are 
the fluid 4--velocities in $\cal M^+$ and $\cal M^-$ respectively. Let 
${}^+u^\mu$ be the same vector as ${}^-u^\mu$ but calculated on 
the plus side of $\cal N$. We then compare (on $\cal N$) ${}^+v^\mu$ 
with ${}^+u^\mu$ and if they are equal then the fluid 4--velocity is 
continuous across $\cal N$. To do this we utilise the tangent basis 
vectors $e_{(a)}=\partial /\xi ^a$ (with $\xi ^a =(r, \phi , z)$ in this 
case) introduced at the beginning of section 2. Then ${}^+u^\mu$ is the 
same vector as ${}^-u^\mu$ if
\begin{equation}\label{4.7}
[u\cdot u]=\left [u\cdot e_{(a)}\right ]=0\ .
\end{equation}
These are the same conditions (\ref{2.4}) that a transversal $N$ 
on $\cal N$ has to satisfy and indeed $u$ can be used as a 
transversal if desired. The four conditions (\ref{4.7}) determine 
${}^+u^\mu$ uniquely and for the example we are considering we 
obtain ${}^+u^\mu =(1, 0, 0, 0)$. Hence ${}^+u^\mu ={}^+v^\mu$ 
and the fluid 4--velocity is continuous across $\cal N$. Now 
using the theory outlined in section 2 above we find that 
$\gamma _{\mu\nu}=0$ except for (quoting the non--vanishing 
components of $\gamma _{\mu\nu}$ in the coordinate system $\{x^\mu _-\}$ 
(say))
\begin{equation}\label{4.8}
\gamma _{11}=\frac{(\beta -2)}{\beta}\,
t^{\frac{(2-\beta )}{2\beta}}\ ,
\hspace{0.3cm}\gamma _{22}=\frac{9\beta\,(\beta -2)}{(3\beta -2)^2}\,
t^{\frac{(9\beta -2)}{6\beta}}\ .
\end{equation}
With $n_-^\mu=(t^{(-3\beta +2)/6\beta}, t^{-(3\beta +2)/6\beta}, 0, 0)$ 
and $m_-^\mu=2^{-1/2}\,t^{-2/3\beta}\,(0, 0, i\,r^{-1}, 1)$ we find 
that
\begin{equation}\label{4.9}
\gamma _\mu=\gamma _{\mu\nu}\,n^\nu _-=\delta ^1_\mu\,\frac{(\beta -2)}
{\beta}\,t^{(2-3\beta )/3\beta}\ ,
\end{equation}
\begin{equation}\label{4.10}
\gamma ^{\dagger}=\gamma _{\mu}\,n^\mu _-=\frac{(\beta -2)}{\beta }\,
t^{(2-9\beta )/6\beta}\ ,
\end{equation}
\begin{equation}\label{4.11}
\gamma _{\mu\nu}\,\bar m^{\mu}\,\bar m^{\nu}=-\frac{(\beta -2)}
{2\beta}\,t^{-(3\beta +2)/6\beta}\ ,
\end{equation}
where in (\ref{4.11}) we have written $r$ in terms of $t$ following from 
(\ref{4.5}). Comparison now with (\ref{3.22}) shows that $\hat\Psi _3=0$ 
but $\hat\Psi _2\neq 0$ and $\hat\Psi _4\neq 0$. 
In addition we find that the vector field $\lambda$ on $\cal N$ 
introduced in (\ref{2.17}) vanishes, as does $U_{\alpha}=\gamma _{\alpha\beta}\,
{}^-u^{\beta}$. We see from (\ref{4.9})--(\ref{4.11}) that 
the geometry induced on $\cal N$ is a Type I geometry and 
that $\cal N$ is the history of both an impulsive gravitational wave 
and a light--like shell.

In the general case of a Type I induced geometry on $\cal N$ we notice 
that the equations following from the Bianchi identities (\ref{3.16})--
(\ref{3.18}) are all algebraic relations between some components of 
$\gamma _{\mu\nu}$ and some of the jumps in the energy--momentum--stress 
tensors and the Weyl tensors of $\cal M^+$ and $\cal M^-$ across $\cal N$. 
The further consequences of the Bianchi identities when the induced 
geometry is Type II or III (equations (\ref{3.7}), (\ref{3.8}), 
(\ref{3.19})--(\ref{3.21})) can all be viewed as propagation equations 
for components of $\gamma _{\mu\nu}$ along the generators of $\cal N$ 
(derivatives along the generators being indicated by a prime). This 
is consistent because the Type I geometry by itself excludes the 
possibility of a unique parameter being assigned to the geodesic 
generators of $\cal N$ on both the plus and minus sides and hence 
unique propagation equations along these generators of quantities 
defined on $\cal N$ cannot exist.

We emphasise the algebraic nature of the Bianchi identities in 
the case of a Type I geometry by stating the following:
\vfill\eject

\noindent
Lemma 2:

\noindent
{\it If the geometry induced on $\cal N$ is Type I then
\vskip 1truepc\noindent
(a) If $\rho\neq 0$ and/or $\sigma\neq 0$, $\hat\Psi _2$ satisfies
\begin{eqnarray}
\rho\,\hat\Psi_2&=&\frac{4\pi}{3}\,\left [T_{\mu\nu}\,n^\mu\,n^\nu\right]\ ,\\
\sigma\,\hat\Psi_2&=&\frac{1}{3}\,\left [\Psi _0\right ]\ ;
\end{eqnarray}
\vskip 1truepc
\noindent
(b) If $\rho ^2\neq |\sigma |^2\neq 0$, $\hat\Psi _3$ is given by
\begin{equation}
\left [\Psi _1\right ]-4\,\pi\,\left [T_{\mu\nu}m^\mu\,n^\nu\right ]
=-\hat\Psi _3^*\,\rho +\hat\Psi _3\,\sigma\ ,
\end{equation}
and its complex conjugate (here $\hat\Psi _3^*$ is the 
complex conjugate of $\hat\Psi _3$)}. 
\vskip 1truepc\noindent
We note again that 
$\rho , \sigma$ are intrinsic to $\cal N$ ($[\rho ]=0
=[\sigma ]$) for a Type I geometry. For the cosmological example 
given above the expansion $\rho$ and shear $\sigma$ of the 
generators of $\cal N$ are given by
\begin{equation}\label{4.11'}
\rho =\frac{1}{r}\,\left (\frac{3\beta +2}{3\beta -2}\right )\hspace{0.25cm}{\rm and}
\hspace{0.25cm}\sigma =\frac{1}{\sqrt{2}\,r}\ ,
\end{equation}
while $\left [T_{\mu\nu}\,n^\mu\,n^\mu\right ]=s^2\,[\mu +p]$, 
with $s$ defined after (\ref{2.18}), and since on $\cal N$ the 
continuous 4--velocity $u$ is orthogonal to the complex null 
vector $m$ tangent to $\cal N$, $\left [T_{\mu\nu}\,m^\mu\,n^\nu\right ]
=0$ and one can readily verify that the algebraic equations in 
Lemma 2 are satisfied.

The richest induced geometry is of course Type III and in this 
case we can, with additional assumptions, deduce from the Bianchi 
identities some interesting conclusions which we summarise in the 
following:
\vskip 1truepc\noindent
Lemma 3:

\noindent
{\it If the geometry on $\cal N$ is Type III and if 
$\cal M^\pm$ are vacuum space--times then $\left [\Psi _0\right ]=
\left [\Psi _1\right ]=0$,
\begin{equation}
\left [\Psi _2\right ]=\sigma\,\hat\Psi _4-4\,\pi\,\eta\,\rho\,f\ ,
\end{equation}
and thus if $\left [\Psi _2\right ]=0$ and $\hat\Psi _4\neq 0$ then
\vskip 1truepc
\noindent (1) $\sigma =0$ and $\rho\neq 0\ \ \Rightarrow\ \ f=0$\ ,
\vskip 1truepc
\noindent (2) $\sigma =0$ and $\rho =0\ \ \Rightarrow\ \ f\neq 0$ is 
possible\ ,
\vskip 1truepc
\noindent (3) $\sigma\neq 0\ \ \Rightarrow\ \ \rho\neq 0$ and 
$f\neq 0$,
\vskip 1truepc\noindent
where the surface stress--energy tensor of the light--like shell 
now has the form $S_{\alpha\beta}=\eta\,f\,n_{\alpha}\,n_{\beta}$}.
\vskip 2truepc
We first note that part (1) of Lemma 3 explains the 
`miracle' whereby the Penrose spherical impulsive 
wave propagating through flat space--time \cite{P} 
{\it automatically} satisfies the vacuum field equations. 
The history $\cal N$ of the signal in this case is 
a future null--cone which is a shear--free ($\sigma =0$) 
expanding ($\rho\neq 0$) null hypersurface. The induced 
geometry is Type III and thus by Lemma 3(1) 
the surface stress/energy tensor $S_{\alpha\beta}$ 
must vanish ($f=0$). An example of part (1) of Lemma 3 
in which $\cal M^\pm$ are not flat is provided by taking 
$\cal M^\pm$ to be two Petrov Type III Robinson--Trautman 
\cite{RT} vacuum space--times with line--elements of 
the form
\begin{equation}\label{4.12}
ds^2_{\pm}=-2r^2_{\pm}p^{-2}_{\pm}d\zeta _{\pm}d\bar\zeta _{\pm}+2du\,
dr_{\pm}+K_{\pm}
\,du^2\ ,
\end{equation}
with $p_{\pm}=p_{\pm}(\zeta _{\pm}, \bar\zeta _{\pm})$ and
\begin{equation}\label{4.13}
K_{\pm}=\Delta _{\pm}{\rm log}\,p_{\pm}\hspace{0.5cm},\hspace{0.5cm}\Delta _{\pm}K_{\pm}=0\ ,
\end{equation}
where $\Delta _{\pm}=2p^2_{\pm}\partial ^2/
\partial\zeta _{\pm}\partial\bar\zeta _{\pm}$. These 
two space--times are joined together on the shear--free, 
expanding null hypersurface $\cal N$ with equation 
$u=0$, with (\ref{2.1}) satisfied if
\begin{equation}\label{4.14}
\zeta _+=h(\zeta _-)\hspace{0.5cm}
{\rm and}\hspace{0.5cm}r_+=F(\zeta _-, \bar\zeta _-)\,r_- \ ,
\end{equation}
where $h$ is an analytic function of $\zeta _-$ and 
$F(\zeta _-, \bar\zeta _-)=p_+/(|h'|p_-)$. In coordinates 
labelled $x^\mu _-=(\zeta _-, \bar\zeta _-, r_-, u)$ 
we find that $\gamma _{\mu\nu}=0$ except for 
$\gamma _{11}$ and $\gamma _{22}=\bar\gamma _{11}$
with 
\begin{equation}\label{4.15}
\gamma _{11}=-2r\frac{F'}{F}\frac{\partial}{\partial\zeta}
{\rm log}\left (F'\,p^2_-\right )\ ,
\end{equation}
where $F'=\partial F/\partial\zeta _-$. Thus the induced 
geometry is Type III, there is no surface stress/energy 
tensor on $\cal N$ and, since $\hat\Psi _4=\frac{1}{2}\gamma _{11}\,r^{-2}_-
p_-^2\neq 0$, $\cal N$ is the history of an impulsive 
gravitational wave.

Part (2) of Lemma 3 shows that if $\cal N$ is a 
null hyper{\it plane} (with generators having 
vanishing shear and expansion) and if the matching 
of $\cal M^+$ and $\cal M^-$ on $\cal N$ satisfying 
(\ref{2.1}) is such that the induced geometry is 
Type III then $\cal N$ can be the history of a plane 
impulsive gravitational wave and/or a plane 
light--like shell of matter. For example take $\cal M^+$ to be 
a pp--wave space--time with line--element 
\begin{equation}\label{4.16}
ds_+^2=dx^2_++dy^2_++2du\,dv_++H(x_+, y_+, u)\,du^2\ ,
\end{equation}
with $H_{x_+x_+}+H_{y_+y_+}=0$ (subscripts here 
denoting partial derivatives). Take $\cal M^-$ to be flat space--
time with line--element
\begin{equation}\label{4.17}
ds^2_-=dx^2+dy^2+2du\,dv\ .\end{equation}
Now match $\cal M^+$ to $\cal M^-$ on the null 
hyperplane $\cal N$ $(u=0)$ with \cite{P}
\begin{equation}\label{4.18}
x_+=x\ ,\hspace{0.5cm}y_+=y\ ,
\hspace{0.5cm}v_+=v+h(x, y)\ ,
\end{equation}
to ensure that (\ref{2.1}) is satisfied. Using 
the theory of sections 2 and 3 above with 
$x^\mu _-=(\xi ^a, u)$ with $\xi ^a=(x, y, v)$ we 
find that $\gamma _{\mu 4}=0$ and otherwise 
$\gamma _{ab}=-h_{,ab}$. Thus with $n_-^\mu=\delta 
^\mu _3$ we have $\gamma _\mu =\gamma _{\mu 3}=0$ 
and so the geometry induced on $\cal N$ is Type III. 
We also find that
\begin{equation}\label{4.19}
8\pi\,f=-\frac{1}{2}\,\gamma =h_{xx}+h_{yy}\ ,
\end{equation}
and
\begin{equation}\label{4.20}
\hat\Psi _4=-\frac{1}{2}\,(h_{xx}-h_{yy})+i\,h_{xy}\ .
\end{equation}
This shows explicitly that a light--like shell and a 
plane impulsive wave can co--exist, each with history $\cal N$.

A simple example of part (3) of Lemma 3 is a cylindrical 
fronted light--like signal with history $\cal N$ in 
flat space--time. Thus $\cal M^\pm$ have line--elements 
\begin{equation}\label{4.21}
ds_{\pm}^2=(u+v_{\pm})^2d\phi ^2_{\pm}+dz^2_{\pm}
+2du\,dv_{\pm}\ .\end{equation}
Now $\cal N$ $(u=0)$ is a null hypersurface 
generated by shearing null geodesics ($\sigma\neq 0$). 
We match the induced metrics on $\cal N$ with 
\begin{equation}\label{4.21}
\phi _+=q(\phi )\ ,\hspace{0.5cm}z_+=z\ ,\hspace{0.5cm}
v_+=v/q'\ ,\end{equation}
with $q'=dq /d\phi$. In coordinates $x^\mu _-=(\phi , z, v, u)
$ we find that $\gamma _{\mu\nu}=0$ except for 
\begin{equation}\label{4.22}
\gamma _{11}=2v\,\left\{\frac{q''}{q'}-\frac{3}{2}\,
\left (\frac{q''}{q'}\right )^2+q'^2-1\right\}\ .
\end{equation}
Thus with $n_-^\mu=\delta ^\mu _3$ we see that 
$\gamma _\mu =0$ and the induced geometry on 
$\cal N$ is Type III. The shear $\sigma$ and expansion 
$\rho$ of the null geodesic generators of $\cal N$ 
satisfy
\begin{equation}\label{4.23}
\rho =\sigma =\frac{1}{2v}\ ,\end{equation}
while
\begin{equation}\label{4.24}
4\pi\,f=\hat\Psi _4=-\frac{1}{4v^2}\,\gamma _{11}\ .
\end{equation}
Thus in general $f\neq 0$ and a shell and impulsive wave 
co--exist. We see that no signal exists with history 
$\cal N$ if and only if $\gamma _{11}=0$. It is interesting to 
note that if no signal exists on $\cal N$ and the isometric transformations 
preserving this state form a {\it group} then they are given by  
(\ref{4.21}) with $q(\phi )=\phi +c$, and $c={\rm constant}$. 
There also exist other disconnected isometric 
transformations of $\cal N$, of the form (\ref{4.21}), 
when no signal exists on $\cal N$, but these transformations 
do not form a group.
\vskip 1truepc
A corresponding Lemma to Lemma 3 which has applications to 
light--like signals propagating through a cosmic fluid is:
\vskip 1truepc 
Lemma 4:

\noindent
{\it If the geometry on $\cal N$ is Type III and if 
$\cal M^\pm$ are perfect fluid space--times with $u$ continuous 
across $\cal N$ then $\left [\Psi _0\right ]=\left [\Psi _1\right ]
=\left [\mu +p\right ]=0$ and }
\begin{equation}
\left [\Psi _2\right ]-\frac{4\pi}{3}\,\left [\mu\right ]
=\sigma\,\hat\Psi _4-4\,\pi\,\eta\,\rho\,f\ ,
\end{equation}
{\it and thus if $\left [\Psi _2\right ]=\frac{4\pi}{3}\,
\left [\mu\right ]$ and $\hat\Psi_4\neq 0$ then the deductions are the same as 
(1) --(3) of Lemma 3}.
\vskip 1truepc
\noindent
The special case of the de Sitter universe is obtained by putting 
$-8\pi\,p=8\pi\,\mu=\Lambda$, where $\Lambda$ is the cosmological 
constant. We will confine our observations on Lemma 4 to the 
de Sitter case leaving further applications of Lemma 4 to another 
occasion.
\vskip 1truepc
To illustrate (4.31) of Lemma 4 we let 
$\cal M^{\pm}$ both be de Sitter universes (with different 
cosmological constants $\Lambda _{\pm}$) having line--elements 
\begin{equation}\label{4.25}
ds^2_{\pm}=\frac{2v^2_{\pm}d\zeta _{\pm}d\bar\zeta _{\pm}
+2du_{\pm}dv_{\pm}}{\left (1+\frac{1}{6}\Lambda _{\pm}u_{\pm}v_{\pm}
\right )^2}\ .\end{equation}
Here $\cal N$ ($u_+=u_-=0$) is a future null--cone generated by 
expanding ($\rho\neq 0$) shear--free ($\sigma =0$) null geodesics. 
We match $\cal M^{\pm}$ on $\cal N$ with a Penrose \cite{P} warp 
\begin{equation}\label{4.26}
\zeta _+=h(\zeta _-)\ ,\hspace{0.5cm}v_+=\frac{v_-}{|h'|}\ ,
\end{equation}
where $h$ is an analytic function of $\zeta _-$ and $h'=dh/d\zeta _-$. 
Now the induced geometry on $\cal N$ is Type III. In general 
$\hat\Psi _4=-\chi /2v\neq 0$ with
\begin{equation}\label{4.27'}
\chi =\frac{h'''}{h'}-\frac{3}{2}\,\left (\frac{h''}{h'}
\right )^2
\ ,
\end{equation}
and 
\begin{equation}\label{4.27}
8\pi\,\rho\,f=\frac{1}{3}\,\left [\Lambda\right ]\ .\end{equation} 
This is the form taken by (4.31) for this example since now ${}^{\pm}
\Psi _A=0$ for $A=0, 1, 2, 3, 4$, $8\pi\,[\mu ]=\left [\Lambda\right ],\ 
\eta =+1$ and $\sigma =0$.
Thus if $\left [\Lambda\right ]=0$ then since $\rho =1/v\neq 0$ we must have $f=0$ and so $\cal N$ is the 
history of an impulsive gravitational wave \cite{PH}.

Finally as an illustration of conclusion (2) of Lemma 4 
we consider $\cal M^+$ to be a Schwarzschild space--
time with line--element in Kruskal form
\begin{equation}\label{4.28}
ds^2=r^2(d\theta ^2+\sin ^2\theta\,d\phi ^2)
-\frac{64m^3}{r}\,{\rm e}^{\left (1-r/2m\right )}\,dU\,dV\ ,
\end{equation}
with $r=r\left (UV\right )$ given by
\begin{equation}\label{4.29}
\left (\frac{r}{2m}-1\right )\,{\rm e}^{\left (\frac{r}{2m}-1\right )}=
2U\,V\ 
,\end{equation}
and we take $\cal M^-$ to be de Sitter space--time 
(with $\Lambda >0$) with line--element 
\begin{equation}\label{4.30}
ds^2=r^2(d\theta ^2+\sin ^2\theta\,d\phi ^2)-2
\frac{(1+\lambda\,r)^2}{\lambda ^2}\,dU\,dV\ ,
\end{equation}
where $\lambda ^2=\Lambda /3$ and $r=r\left (UV\right )$ 
is given by
\begin{equation}\label{4.31}
\frac{1-\lambda\,r}{1+\lambda\,r}=2U\,V\ .\end{equation}
These match (cf. \cite{BI}) on the horizon $\cal N$ $\left (U=0\right )$ 
if $2m\,\lambda =1$. We have rescaled one of the null 
coordinates in (\ref{4.28}) and (\ref{4.29}) to 
make the metric tensors given via the line--elements 
(\ref{4.28}) and (\ref{4.30}) continuous across $\cal N$. 
The horizon $U=0$ is a null hyperplane generated by shear--free ($\sigma 
=0$), expansion--free ($\rho =0$) null geodesics. In the 
continuous coordinates $\left (U, V, \theta , \phi\right )$ 
above we find using (2.12) that $\gamma _{\mu\nu}=0$ 
except $\gamma _{22}=\gamma _{11}\,\sin ^2\theta =-
3V\,\sin ^2\theta /4m^2$ and $f=3V/32\pi m^2\neq 0$. If the situation above is reversed 
and $\cal M^+$ is de Sitter space--time (with $\Lambda >0$) 
and $\cal M^-$ is Schwarzschild space--time then $\gamma _{\mu\nu}
=0$ except for $\gamma _{22}=\gamma _{11}\,\sin ^2\theta =
3V\,\sin ^2\theta /4m^2$ and $f=-3V/32\pi m^2$. In either 
case the induced geometry is type III. The equation $\left [\Psi 
_2\right ]=\frac{4\pi}{3}\,\left [\mu\right ]$ becomes 
$\frac{1}{4m^2}=\lambda ^2$. There is no gravitational 
wave present ($\hat\Psi _4 =0$) and $\cal N$ 
is the history of a light--like shell.
\vfill\eject\setcounter{equation}{0} 
\section{Discussion}\indent
The Lemmas that we have established above 
fall into two different categories. Lemma 1 concerns 
the interaction between a null shell and/or a wave 
and any time--like congruence with a continuous tangent 
vector at the intersection with $\cal N$. It shows the close 
relationship existing between the presence 
of a wave ($\hat\Psi _4\neq 0$) and the shear of the 
time--like congruence ($\left [\sigma _{\mu\nu}m^\mu\,m^\nu
\right ]\neq 0$). There is a complementary result due to 
Penrose \cite{P} to the effect that for a null geodesic congruence 
crossing $\cal N$ with continuous tangent, a jump in the 
complex shear is necessary for $\cal N$ to be the history 
of an impulsive gravitational wave. On the other hand 
Lemmas 2 -- 4 relate the properties of the null hypersurface 
(embodied in $\rho , \sigma , \hat\Psi _A$) to the 
outside medium (described by ${}^{\pm}T^{\mu\nu}$) and 
the outside geometry (described by ${}^{\pm}\Psi _A$). The different 
jumps $\left [T_{\mu\nu}\,n^\mu\,n^\mu\right ]\ ,
\ \left [T_{\mu\nu}\,m^\mu\,n^\mu\right ]\ ,
\ \left [T_{\mu\nu}\,l^\mu\,l^\mu\right ]$ tell how 
the fluid lines and fluid properties (energy density, 
pressure) are modified by the presence of the light--like signal. 
For instance (3.6) can be written as 
\begin{equation}\label{5.1}
\nabla _\nu\left (\alpha\,S^{\mu\nu}\right )=-\alpha\,
\left [T^{\mu\nu}\,n_{\nu}\right ]\ ,
\end{equation}
and represents an equation of conservation for energy 
and momentum \cite{BI}. For a pure impulsive gravitational 
wave it reduces to $\left [T^{\mu\nu}\,n_\nu\right ]=0$, which 
also holds for a shock wave \cite{WI}, \cite{RG}.

Finally a couple of technical points which have arisen above merit 
discussion. At the beginning of section 3 we 
note that in a local coordinate system $\{x^\mu\}$ covering both 
sides of $\cal N$ the equation of $\cal N$ is $\Phi (x^\mu )=0$ (say) 
and so as normal we can take $n_\mu =\alpha ^{-1}\Phi _{,\mu}$ 
where $\alpha$ is some function defined on $\cal N$. In the passage 
from Type I geometry to Type II geometry the acceleration parameter 
$\kappa$ becomes continuous across $\cal N$. In this case we are 
entitled to put $\alpha =1$ and so make $\kappa$ vanish on $\cal N$. 
However situations can arise in applications with a Type II or 
Type III induced geometry on $\cal N$ in which the 
most natural parameter to use along the generators of $\cal N$ is not 
an affine parameter. Then although $[\kappa ]=0$ we have $\kappa\neq 0$ 
and for this reason we have retained $\kappa$ in equations (3.9), 
(3.21) and (3.22) ($\kappa$ does not appear in (3.8) or (3.20) even 
when non--zero).

In introducing a time--like congruence crossing the history $\cal N$ of 
the light--like signal in section 2 we chose to examine the case 
in which the unit time--like tangent vector field $u^\mu$ is continuous 
across $\cal N$ but may have a jump in its derivative described via 
a vector field $\lambda ^\mu$ introduced in (2.17). As we pointed 
out in section 2 this assumption forbids $\cal N$ being the history 
of a shock wave in the usual sense (a shock wave in a gas with macroscopic 
4--velocity $u^\mu$, for 
example \cite{WI}, \cite{RG}). In this latter case the tangent 
to the congruence would itself jump across $\cal N$. This 
complicates the study of the interaction of the time--like 
congruence with the light--like signal by introducing delta 
functions (singular on $\cal N$) into the kinematical 
variables associated with the congruence and is a 
topic for further study.
\vskip 1truepc\noindent
This research has been financially supported by 
CNRS/Forbairt/MAE.

\end{document}